\documentclass[aps]{revtex4}
\usepackage{graphicx}
\begin{document}
\begin{flushright}
UCRL-JRNL-207489
\end{flushright}
\baselineskip=18pt

\title{\bf  No-core shell model for 48-Ca, 48-Sc and 48-Ti}
\author{ J.P. Vary$^{a,b}$, S. Popescu$^c$, S. Stoica$^c$,  P. Navr\'atil$^b$ \\
\it $^a$ Department of Physics and Astronomy, Iowa State University, \\
\it Ames, IA 500011 \\
\it $^b$ Lawrence Livermore National Laboratory, Livermore, CA 94551,\\
\it $^c$ Horia Hulubei National Institute of Physics and
Nuclear Engineering\\
\it P.O. Box MG-6, 76900 Bucharest-Magurele, Romania \\}
\maketitle
\pagestyle{empty}
\vskip5mm

\noindent
{\bf Abstract}

We report the first no-core shell model results for
$^{48}Ca$, $^{48}Sc$ and $^{48}Ti$ with derived and
modified two-body Hamiltonians.  We use an oscillator basis
with a limited $\hbar\Omega$ range around
$45/A^{1/3}-25/A^{2/3} = 10.5 MeV$
and a limited model space up to $1~\hbar\Omega$.
No single-particle energies are used.
We find that the
charge dependence of the bulk binding energy of eight $A=48$ nuclei is
reasonably described with an effective Hamiltonian derived from the
CD-Bonn interaction while there is an overall underbinding by
about 0.4 MeV/nucleon. However, the resulting spectra exhibit
deficiencies that are anticipated due to: (1) basis space limitations
and/or the absence of effective many-body interactions; and, (2) the
absence of genuine three-nucleon interactions.
We then introduce additive isospin-dependent central terms plus a 
tensor force to our Hamiltonian and achieve accurate binding energies 
and reasonable 
spectra for all three nuclei.
The resulting no-core shell model
opens a path for applications to the 
double-beta ($\beta\beta$) decay process.  

\section{Introduction}

Motivated by the need for precision treatments of nuclear
matrix elements involved in the double-beta ($\beta\beta$) decay process, 
we investigate the {\it ab initio}
no core shell model (NCSM) for the lightest case of such a decay, which 
involves the $^{48}Ca$, $^{48}Sc$ and $^{48}Ti$ nuclei.

Previous efforts have addressed $A=48$ nuclei with
similar goals by solving the shell model of eight nucleons in the pf shell
with a $^{40}Ca$ core.  Chief among these efforts is the work of 
Caurier et al. \cite{[CA90]} who made predictions for the
$2\nu\beta\beta$ decay mode of $^{48}Ca$.  They carried out a full
0$\hbar\Omega$ calculation for the nuclei
involved in this decay, using a KB3 effective interaction \cite{[KB3]}.
Single-particle energies extracted from experiment
were used to model the valence-core interactions.  They obtained a
satisfactory description of the spectroscopy
(energy levels of positive parity, E2 and M1 transitions)
when they used a factor of 0.77 to renormalize the nuclear matrix elements (NME)
of the Gamow-Teller (GT) operator $\sigma\tau_{\pm}$.
This quenching factor accounts for possible missing nucleon
correlations and/or non-nucleonic contributions. 

Similar calculations have been
performed using either slightly different parameters in
the Hamiltonian  \cite{[OH89]} or more severely truncated model spaces
\cite{[ZBR90]}, but the predicted
half-lives differ from the experimental value by a factor 2-3.
Such calculations have also been extended to the
$0\nu\beta\beta$ decay mode of $^{48}Ca$ \cite{[RCN95]}.
The authors used two NN interactions (KB3 and RVJB
\cite{[RIH91]}) and performed calculations using both the full pf shell
model space and another three truncated spaces. They concluded that in
this case the results are not significantly dependent on the truncation
as long as the 2p-2h correlations are included.

Caurier et al. \cite{[CA94]} also performed a systematic
calculation of the spectroscopic properties
(positive parity energy levels, B(E2), M1, quadrupole moments, GT
strengths) of several $A=48$ isotopes: $^{48}Ca$, $^{48}Sc$, $^{48}Ti$,
$^{48}V$, $^{48}Cr$ and $^{48}Mn$. The model space consisted again of
8-nucleons in
the full pf shell and they used a minimally monopole modified
realistic KB3 interaction along with experimental single-particle
energies. Their results reproduced well the existing data with very few
parameters. The calculated GT strengths are in agreement with
experiment using the standard quenching factor mentioned above and
display a fine structure which indicates that fragmentation renders
much of the GT strength unobservable.

Alternatively, in ref. \cite{[POV95]} the authors
performed a full pf shell model calculation of the $2\nu\beta\beta$ decay
rate of $^{48}Ca$ using a Hamiltonian given as a set of 195 two-body NME. 
They examined how modifications in the Hamiltonian  
influence the value of the decay rate and concluded
that with an appropiate change in the pairing interaction one may obtain
satisfactory agreement with experiment without renormalizing the GT operators.
In addition, they found that  making the quadrupole force more attractive is
equivalent with making the pairing weaker. 

Other works performing a shell model calculation of the $2\nu\beta\beta$ NME
emphasize the role of the nuclear shell structure, which gives rise to a
concentration or fragmentation of the decay rate components over the
intermediate states (\cite{[NAK98]}) and the important role of the spin-orbit 
term, compared with the other terms of a schematic potential that also
included central and tensor terms (\cite{[CDN01]}).

Finally, we mention that more details on the capabilities of 
the Shell Model to describe complex manifestations of nuclear dynamics can be 
found in a recent review written by Caurier et al.(\cite{Caurier05}).
 
As an alternative approach, we investigate
the NCSM for these $A=48$ nuclei. This is the first time
such calculations are reported for nuclei with $A \ge 17$.
Our initial goal is to demonstrate the limited feasibility of such NCSM
calculations in this region.  We will display the current status of the
NCSM and, with phenomenology, also demonstrate how far we still
must go before addressing fundamental
processes such as $\beta\beta$ decay rates.  Due to the
considerable gap between present NCSM theory and the best
fit results, we defer evaluations of decay rates to later efforts.

The main ingredients of our approach are the following:

1) We adopt the NCSM approach and approximate the full $H_{eff}$
with a two-body cluster truncation.  In solving the $A=48$ systems,
all nucleons are treated with the same two-body Hamiltonian
derived from a realistic NN interaction including Coulomb interaction
between proton pairs.  There are no single-particle
energies involved and the eigenenergies are the total binding
energies.

2) We work in a neutron-proton basis so full isospin mixing is allowed
including isospin mixing arising from the bare NN interaction and
those induced within Heff itself.

3) For 48-Ca we evaluate both the positive (0hw) and negative parity
(1hw) spectra.

4) Our solutions are free of spurious center of mass motion effects.

5) Our wavefunctions exactly obey the Ikeda sum rule \cite{Ikeda}.

6) We provide a baseline for further improvements to Heff such as the
inclusion of real and effective three-body forces.

7) We show that by introducing of a few phenomenological terms the 
description of both the BE/A and the low-lying spectra for all three nuclei 
improve considerably.
 
Of course the present work has a significant drawback. Due to
the limited model space and the neglect of real and effective
three-body interactions,
we must resort to additive phenomenological terms to obtain high
quality description of selected experimental data. The dependence
on the parameters introduced, including the basis space parameters
$N_{max}$ and $\hbar\Omega$, as well as dependence on the forms and
strengths of the additive potential terms, severely limit the
predictive power of our present approach.
On the other hand, the descriptions achieved with our
initial choice of additive terms provides insight into the
deficiencies of our current Heff in the $0\hbar\Omega$ and
$1\hbar\Omega$ model spaces.

The paper is organized as follows. In section 2 we give a brief
description of the NCSM formalism. In section 3 we present
our results and analysis and section 4 is devoted
to conclusions and perspectives.

\section{No core shell model (NCSM) formalism}

In order to take a step towards increased predictive power, we adapt
the  {\it ab initio} NCSM as we
will briefly outline here. However, due to
computational limitations translating into model space limitations,
we expect and find that the spectroscopic results differ 
significantly from experiment.
Since previous investigations summarized above
have indicated the importance of obtaining
good spectroscopic descriptions in order to minimize uncertainties in
predictions of key experimental observables, we will introduce additive
phenomenological corrections sufficient to fit experimental spectra.

\subsection{Basic NCSM and Effective Hamiltonian}

The NCSM~\cite{ZBV}-\cite{NCSM6} is
based on an effective Hamiltonian derived from realistic
``bare'' interactions and acting within a finite Hilbert space.
All $A$-nucleons are treated on an
equal footing.  The approach is demonstrably convergent
to the exact result of the full (infinite) Hilbert space.

Initial investigations used two-body interactions~\cite{ZBV}
based on a G-matrix approach. Later, we implemented
the Lee-Suzuki-Okamoto procedure~\cite{LS80,S82SO83}
to derive two-body and three-body effective interactions
based on realistic NN and NNN interactions.


For pedagogical purposes, we outline the NCSM approach
with NN interactions alone and point the reader to the
literature for the extensions to include NNN interactions.
We begin with the purely intrinsic
Hamiltonian for the $A$-nucleon system, i.e.,
\begin{equation}\label{ham}
H_A= T_{\rm rel} + {\cal V} =
\frac{1}{A}\sum_{i<j}^A \frac{(\vec{p}_i-\vec{p}_j)^2}{2m}
+ \sum_{i<j=1}^A V_{\rm N}(\vec{r}_i-\vec{r}_j) \; ,
\end{equation}
where $m$ is the nucleon mass and $V_{\rm N}(\vec{r}_i-\vec{r}_j)$,
the NN interaction, with both strong and electromagnetic components.
Note the absence of a phenomenological single-particle potential.
We may use either coordinate-space
NN potentials, such as the Argonne potentials \cite{GFMC}
or momentum-space dependent
NN potentials, such as the CD-Bonn \cite{Machl} which we select
for the present investigation.

Next, we add the center-of-mass HO Hamiltonian
to the Hamiltonian (\ref{ham})
$H_{\rm CM}=T_{\rm CM}+ U_{\rm CM}$, where
$U_{\rm CM}=\frac{1}{2}Am\Omega^2 \vec{R}^2$,
$\vec{R}=\frac{1}{A}\sum_{i=1}^{A}\vec{r}_i$.
In the full Hilbert space the added  $H_{\rm CM}$ term
has no influence on the intrinsic properties.
However, when we introduce our cluster approximation below,
the added  $H_{\rm CM}$ term
facilitates convergence to exact results
with increasing basis size.
The modified Hamiltonian, with a pseudo-dependence on the HO
frequency $\Omega$, can be cast into the form
\begin{equation}\label{hamomega}
H_A^\Omega= H_A + H_{\rm CM}=\sum_{i=1}^A \left[ \frac{\vec{p}_i^2}{2m}
+\frac{1}{2}m\Omega^2 \vec{r}^2_i
\right] + \sum_{i<j=1}^A \left[ V_{\rm N}(\vec{r}_i-\vec{r}_j)
-\frac{m\Omega^2}{2A}
(\vec{r}_i-\vec{r}_j)^2
\right] \; .
\end{equation}

In the spirit of Da Providencia and Shakin \cite{PS64} and Lee,
Suzuki and Okamoto \cite{LS80,S82SO83}, we introduce a
unitary transformation, which is able to accommodate
the short-range two-body correlations in a nucleus,
by choosing an antihermitian operator $S$, acting only on
intrinsic coordinates, such that
\begin{equation}\label{UMOAtrans}
{\cal H} = e^{-S} H_A^\Omega e^{S} \; .
\end{equation}
In our approach, $S$ is determined by the requirements that ${\cal H}$
and $H_A^\Omega$ have the same symmetries and eigenspectra over
the subspace ${\cal K}$ of the full Hilbert space.
In general, both S and the transformed Hamiltonian are
$A$-body operators.
Our simplest, non-trivial approximation to ${\cal H}$ is to develop
a two-body $(a=2)$ effective Hamiltonian, where the upper bound
of the summations ``$A$'' is replaced by ``$a$'', but
the coefficients remain unchanged.
The next improvement is to develop
a three-body effective Hamiltonian, $(a=3)$.   This approach consists
then
of an approximation to a particular level of clustering with $a \leq
A$.
\begin{equation}\label{UMOAexpan}
{\cal H} = {\cal H}^{(1)} + {\cal H}^{(a)} =  \sum_{i=1}^{A} h_i +
\frac{{A \choose 2}}{{A \choose a}{a \choose 2}}
\sum_{i_{1}<i_{2}< \ldots <i_{a}}^{A}\tilde{V}_{i_{1}i_{2} \ldots
i_{a}} \; ,
\end{equation}
with
\begin{equation}\label{UMOAexplterms}
\tilde{V}_{12 \ldots a} = e^{-S^{(a)}}H^{\Omega}_{a}e^{S^{(a)}}
- \sum_{i=1}^a h_i \; , \\
\end{equation}
and $S^{(a)}$ is an $a$-body operator;$H^{\Omega}_{a} = h_1+h_2+h_3+
\ldots
+h_{a}+V_{a}$,
and $V_{a} = \sum_{i<j}^{a} V_{ij}$.
Note that there is no sum over ``$a$'' in Eq. (\ref{UMOAexpan}) so
there is no coupling between clusters in this approach.
Also, we adopt the HO basis states that are eigenstates of
the one-body Hamiltonian $\sum_{i=1}^A h_i$.

If the full Hilbert space is divided into a finite model space
(``P-space'') and a complementary infinite space (``Q-space''), using
the projectors $P$ and $Q$ with $P+Q=1$,
it is possible to determine the transformation operator $S_{a}$
from the decoupling condition
\begin{equation}\label{UMOAdecoupl}
Q_{a} e^{-S^{(a)}}H^{\Omega}_{a}e^{S^{(a)}} P_{a} = 0
  \; ,
\end{equation}
and the simultaneous restrictions $P_a S^{(a)} P_a = Q_a S^{(a)} Q_a
=0$.
Note that $a$-nucleon-state projectors ($P_a, Q_a$)
appear in Eq. (\ref{UMOAdecoupl}). Their definitions follow from the
definitions of the $A$-nucleon projectors $P$, $Q$.
The unitary transformation and decoupling
condition, introduced by Suzuki and
Okamoto and  referred to as the unitary-model-operator approach (UMOA)
\cite{UMOA},
has a solution that can be expressed in the following form
\begin{equation}\label{UMOAsol}
S^{(a)}  = {\rm arctanh}(\omega-\omega^\dagger)    \; ,
\end{equation}
with the operator $\omega$ satisfying $\omega=Q_a\omega P_a$,
and solving its own decoupling equation,
\begin{equation}\label{decoupl}
Q_a e^{-\omega}H^{\Omega}_{a}e^\omega P_a = 0 \; .
\end{equation}
Let us also note that
$\bar{H}_{\rm a-eff}=P_{a} e^{-S^{(a)}}H^{\Omega}_{a}e^{S^{(a)}}
  P_{a}$
leads to the relation
\begin{equation}\label{exhermeff}
\bar{H}_{\rm a-eff}
= (P_{a}+\omega^\dagger\omega )^{-1/2}
(P_{a}+P_{a}\omega^\dagger Q_{a})H_{a}^\Omega (Q_{a}\omega P_{a}
+P_{a})(P_{a}+\omega^\dagger\omega )^{-1/2}  \; .
\end{equation}
Given the eigensolutions,
$H_{a}^\Omega|k\rangle = E_k |k\rangle $,
in the infinite Hilbert space for the cluster,
then the operator $\omega$ can be determined from
\begin{equation}\label{omegasol}
\langle\alpha_Q|\omega|\alpha_P\rangle = \sum_{k \in{\cal K}}
\langle\alpha_Q|k\rangle\langle\tilde{k}|\alpha_P\rangle \; ,
\end{equation}
where we denote by tilde the inverted matrix of
$\langle\alpha_P|k\rangle$, i.e.,
$\sum_{\alpha_P}\langle\tilde{k}|\alpha_P\rangle\langle\alpha_P
|k'\rangle = \delta_{k,k'}$ and
$\sum_k \langle\alpha'_P|\tilde{k}\rangle \langle k|\alpha_P\rangle
= \delta_{\alpha'_P,\alpha_P}$, for $k,k'\in{\cal K}$. In the relation
(\ref{omegasol}),  $|\alpha_P\rangle$ and $|\alpha_Q\rangle$
are the model-space and the Q-space basis states, respectively,
and ${\cal K}$ denotes a set of $d_P$ eigenstates, whose properties
are reproduced in the model space,
with $d_P$ equal to the dimension of the model space.

In practice, the exact (to numerical precision)
solutions for the a=2 cluster are obtained
in basis spaces of several hundred $\hbar\Omega$
in each relative motion NN channel.

We note that in the limit $a \rightarrow A$, we obtain the exact
solutions for $d_P$ states of the full problem for any finite basis
space, with flexibility for the choice of physical states subject to
certain conditions \cite{Viaz01}.  We also note that more compact
methods have recently been introduced to obtain effective interactions
that are equivalent to the methods we use and outline above \cite{Lisetskiy08}.

We define our P-space to consist of all A-particle configurations
in the oscillator basis with oscillator energy less than or equal to some
cutoff value, $N_m = N_{min}+N_{max}$, where $N$ is the sum of $2n+l$
values of the occuppied single-particle states in the configuration.
$N_{min}$ is the minimum value required by the Pauli principle
and equals $84$ for these $A=48$ nuclei. Our P-spaces are equally
described by the excitations allowed through $N_{max}$ which
begins with $0$.  The cluster space, $P_2$, is defined by the range
of 2-body states encountered in the P-space.

Due to our cluster approximation
a dependence of our results on $N_{max}$ and on $\hbar\Omega$ arises.
For a fixed cluster size, the smaller the basis space, the larger the
dependence on $\hbar\Omega$.  The residual $N_{max}$ and
$\hbar\Omega$ dependences can be used to infer the uncertainty
in our results arising from the neglect of effective
many-body interactions.

In light nuclei, the strategy has been to evaluate $H_{eff}$
for each model space leading to a separate $H_{eff}$ for
positive and negative parity states.  As one proceeds to heavier
systems we reason that a better strategy is to
use the same $H_{eff}$ for both positive and negative parities
e.g. use the $1\hbar\Omega$ $H_{eff}$ in both the
$0\hbar\Omega$ and $1\hbar\Omega$ model spaces.
The logic for the revised strategy stems from three considerations:
(1) either strategy will converge to the exact result in
sufficiently large basis spaces; (2) for adjoining spaces in
heavier systems, the predominant sets of pairwise interactions
are in the same configurations with just one pair at a time
shifting to the larger space; and (3) for electromagnetic transitions
between states of opposite parity, the theory of the corresponding 
effective operators will be simplified.  

To elaborate on the second point, we note that the bulk of the binding
should not be altered in proceeding from a
$0\hbar\Omega$ to a $1\hbar\Omega$ model space in A=48,
suggesting the same $H_{eff}$ is preferred. For small model
spaces, such as those investigated here, the revised strategy
improves the splitting between positive and negative parity
states for the physical reason just mentioned.

This revised strategy simplifies our work to evaluate $H_{eff}$ since it is
required only for every other increment in the basis space, such
as $1\hbar\Omega$, $3\hbar\Omega$, $5\hbar\Omega$, etc.,
to evaluate the converging sequence.  A recent work has shown 
the advantages of this revised strategy for improving the converging
sequence in light nuclei - especially the systematic behavior of spectra
at low $N_{max}$ values \cite{Forssen08}.

In order to construct the operator $\omega$ (\ref{omegasol})
we need to select the set of eigenvectors ${\cal K}$.
We select the lowest states obtained in each two-body channel.
It turns out that these states also have the largest overlap
with the model space for the range of $\hbar\Omega$
we have investigated and the P-spaces we select.

We input the effective Hamiltonian, now consisting of a relative
two-body operator and the pure $H_{CM}$ term introduced earlier, into
an m-scheme Lanczos diagonalization process to obtain the P-space
eigenvalues and eigenvectors \cite{MFD}.  At this stage
we also add the term $H_{CM}$ again with a large positive coefficient
to separate the physically interesting states with $0s$ CM motion from
those with excited CM motion.  We retain only the eigenstates with pure
$0s$ CM motion when evaluating observables.  All observables that
are expressible as functions of relative coordinates, such as the rms
radius and radial densities, are then evaluated free of CM motion
effects.

In the case of the $0\hbar\Omega$ model space for $^{48}Ca$,
the neutrons occupy part of the pf shell while the protons fill the
sd shell. Now, for the $1\hbar\Omega$ model space that we adopt for
$^{48}Ca$, we take proton pairs in the Q-space as those with
relative harmonic oscillator states having $(2n+l) \ge 6$.
Similarly, we take neutron pairs
in the $1\hbar\Omega$ Q-space having $(2n+l) \ge 8$.
For the neutron-proton
pairs we take the Heff with pairs in the Q-space of $(2n+l) \ge 7$.
This defines the effective Hamiltonian for both positive and negative
parity states.

It is important to note that we retain this same Hamiltonian
for all the $A=48$ results presented here even though some
have protons in the pf shell.  In so doing, we recognize that
these are additional approximations that we expect
to become less severe in future efforts with enlarged P-spaces.

Since we have chosen separate P-spaces for the neutrons and the
protons, we felt the need to confirm that our treatment of the
spurious CM motion remains valid. We tested this with the
$^{48}Ca$ positive parity spectrum in the
following way. We lowered the Lagrange multiplier of the $H_{CM}$ term
from our conventional value of 10.0 to a value of 2.0.  In the
past this was more than sufficient to reveal any deficiencies
in our treatment of CM motion. In the present case this means that, since
$\hbar\Omega=10 MeV$, our lowest spurious states are around
20 MeV of excitation in $^{48}Ca$
(assuming the CM motion is treated correctly). This is about
as low as we can safely go for a test since we still have significant
separation of the spurious from the non-spurious states.

The test showed that six of the lowest 15 eigenvalues changed
by 1 keV, a change in the 6th significant figure, while the other
9 were unchanged at this precision, indicating at most a change
in the 7th significant figure. This numerically demonstrates
that the CM motion in the NCSM is
accurately treated by the constraint method even when neutrons
and protons occupy different shells, as long as the model space
is defined with a many-body cutoff as we have implemented.

The m-sheme basis dimensionalities are (12022, 139046, and 634744) for
($^{48}Ca$, $^{48}Sc$, and  $^{48}Ti$) respectively in the
$0\hbar\Omega$ model spaces, and 2921360 for $^{48}Ca$
in the $1\hbar\Omega$ model space.  By way of reference,
$^{48}Ca$ in the $2\hbar\Omega$ model space produces an m-scheme matrix
dimension of 214664244.

Alternative renormalization methods, that are independent of
the P-space and independent of the nucleus, have recently been
investigated in light nuclei and shown to significantly soften the
interaction so as to improve convergence rates with increasing 
$N_{max}$ \cite{Bogner08}.  
Each method, in principle, also induces effective many-body 
interactions. The contributions of induced effective 
NNN interactions will soon be evaluated within these methods 
to assure that accurate results are obtained \cite{Furnstahl09}.

Calculations with a new, non-local and soft NN interaction, 
JISP16 \cite{Shirokov07}, have recently been reported 
in light nuclei \cite{Maris09}.  These investigations display
a possible new route for converged calculations without
invoking a renormalization program.  Although convergence
seems assured in light nuclei, it will be a major challenge to
apply this interaction to heavier systems without invoking a
suitable renormalization program and retaining the induced
NNN interactions.

We close our outline on the theoretical framework with the observation
that all observables require the same transformation as implemented on
the Hamiltonian.  To date, we have found rather small effects on the
rms radius operator when we transformed it to a P-space effective
rms operator at the a=2 cluster level \cite{NCSM12}. On the other hand,
substantial renormalization was observed for the kinetic
energy operator when using the $a = 2$ transformation to evaluate
its expectation value \cite{benchmark01}.  More recently, the
underlying physics of effective operator renormalization has 
been explored in detail and major improvements in 
the short-range and/or high momentum transfer properties of nuclei
have been found to be accurately treated by the theory \cite{Stetcu06}.

It is especially noteworthy that our NCSM treatment in the full 
$0\hbar\Omega$ basis space exactly preserves the 
Ikeda sum rule \cite{Ikeda}.  We verified this by explicit calculations
in our applications here.

\subsection{Phenomenological adjustments}

To obtain NCSM spectroscopies fit to the data for these
$A=48$ nuclei by means of additive phenomenological potentials
is a major undertaking.  Hence,
we investigate here minimal approaches to modifying the theoretical
Heff to improve selected spectroscopic properties.
We consider this as a baseline effort for future investigations
in larger model spaces where we believe there will be a
reduced need for phenomenological terms.
Our overall fitting strategy is to emphasize the total
binding energy and the lowest lying excited states.

Inspired by successful modifications found in Ref. \cite{[CA94]},
we first investigated whether a phenomenological S-wave or monopole
interaction supplies the main missing ingredient from our NCSM
realistic effective two-body Hamiltonians.  We chose to add simple
$T=0$ and $T=1$ delta functions and we found that they can produce
greatly improved properties.  However,
we find it necessary to adjust the $T=0$ and $T=1$ strengths for
each nucleus to obtain the good agreement with experimental properties.
Thus, we conclude that, with this approach, six parameters are
needed to obtain reasonable results for the binding energies
and the positive parity spectra of the three nuclei we address.
However, the spectrum of $^{48}Sc$ is still rather poor.

The recent review of the phenomenological shell model 
and the role of the monopole interaction \cite{Caurier05} demonstrates that
the ultimate source of the residual physics we will need is contained in 3-body interaction
 effects - probably a combination of core-polarization and realistic (bare)
 3-body forces.  Our long term goal is to include these additional contributions
 which are natural in our NCSM approach but require next-generation 
 computers.  In order to appreciate the magnitude of the effort needed
 and the potential success of including realistic NN and NNN interactions in large basis
 spaces,  we refer the reader to the recent investigations with Hamiltonians derived
 from chiral effective field theory \cite{Nogga06,Navratil07}.
 
 In addition, more detailed examination of the features of our results (see below) 
 and comparisons with NCSM results in light nuclei indeed indicate that the missing physics
 is tied to larger basis spaces and to realistic NNN interactions.  The fact that a simple
 monopole term in the conventional shell model with a core successfully approximates 
 all this complexity at the 2-body level is remarkable and deserves 
 more extensive investigation.
 
In the hope of obtaining a single NCSM Hamiltonian for the
binding energies and spectra of these three
nuclei, we then explored the utility of two-body central plus
tensor forces added to the {\it ab initio} $H_{eff}$.  We achieve a
reasonable description of a small set of the targeted
properties in these three nuclei by fitting the strengths
and ranges of these three terms.

The specific forms of the finite range central and
tensor potentials we found acceptable are as follows:

\begin{equation}\label{Vcentraltens}
V(r) = V_0 exp(- (r/R)^2)/r^2 + V_1 exp(- (r/R)^2)/r^2
+ V_t S_{12} /r^3
\end{equation}
where the central strengths, $V_0 = -14.40~MeV-fm^2$
and $V_1 = -22.61~MeV-fm^2$
with $R = 1.5 fm$, the
tensor strength $V_t = -52.22~MeV-fm^3$, and $S_{12}$ is
the conventional tensor operator.




\section{Results and Discussion}

We begin with a survey of the ground state eigenenergies of
$A=48$ nuclei presented in Fig. 1.
The experimental values and
extrapolations based on systematics \cite {Audi2003} are
presented as square points and portray the valley of stability
with $^{48}Ti$ being the most stable $A=48$ system. The
upper set of results (round dots)
are those obtained with the {\it ab initio} NCSM as outlined above
with the CD-Bonn interaction \cite {Machl} and $\hbar\Omega=10~MeV$, 
a typical choice for this region.
Note that the trend of the even-even and odd-odd nuclear
binding energies matches reasonably well with experiment except
that theory consistently underbinds by about 20 MeV
(0.4 MeV/nucleon).  In other words, except for this
underbinding, the {\it ab initio} NCSM already predicts some subtle
features of the valley of stability.

\begin{figure}[ht]
\label{fig:Aeq48_BE}
\includegraphics[width=\textwidth,clip]{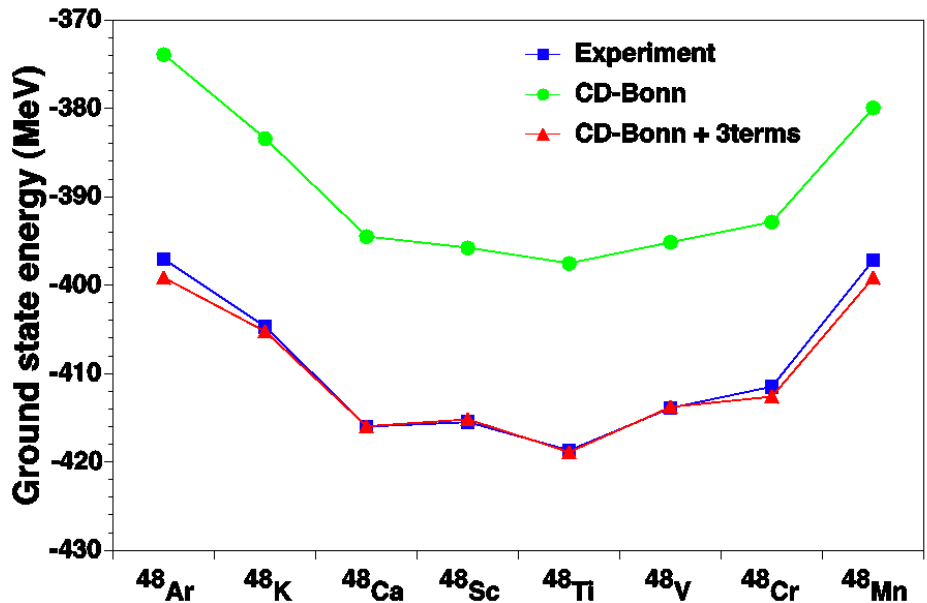}
\caption{(Color online) Ground state energies in MeV of
$A=48$ nuclei.  At the extremes of the valley of stability, these energies
are determined by systematics and the label "experiment" is to be
understood in that context.  The {\it ab initio} NCSM results labeled "CD-Bonn" 
\cite {Machl} are obtained with $H_{eff}$ evaluated for
 the $1\hbar\Omega$ model space, $\hbar\Omega=10~MeV$ and isospin breaking
in the P-space, as described in the text.  The same
$H_{eff}$ with added gaussian central terms and a tensor force is used for the
results labeled "CD-Bonn + 3 terms". }
\end{figure}

While all even-even nuclei obtained here in the {\it ab initio} NCSM have
the correct $J^{\pi}=0^+$ ground state spin and parity, the
odd-odd nuclei generally have the incorrect ground state spin.

The point proton (neutron) rms radii are 3.51, 3.55, and 3.59 fm
(3.75, 3.73, and 3.71 fm) for $^{48}Ca$, $^{48}Sc$ and $^{48}Ti$
respectively, when evaluated with the bare operator in
the intrinsic coordinate system.  Of course, due to the limited
model space, these results are insensitive to configuration
mixing and are controlled by the choice of $\hbar\Omega$.

Increasing $\hbar\Omega$ leads to increased binding (and decreased
rms radii) in this application of the {\it ab initio} NCSM
($\hbar\Omega=10.5MeV$ would produce a good fit to the binding alone)
but fails to improve the errors in ground state spins and
other deficiencies in the spectral properties described below
in more detail.

The overall binding energy picture is considerably improved
with the phenomenological
additions described above.  These additive terms were fit
by hand to the ground state energies of $^{48}Ca$, $^{48}Sc$
and $^{48}Ti$ as well as the first excited positive
and negative parity states in $^{48}Ca$.  This limited amount of
data under-constrains the fit and alternative parameterizations
of the additive terms would yield equivalent fits to these
limited data.  Our approach here was to cease fitting when the first
successful fit was obtained.  Hence, all other properties of these
$A=48$ nuclei, including the ground state energies of the
remaining 5 nuclei presented in Fig. 1,
constitute predictions of
this model.  With the additive terms, the ground state spins and parities
for the eight nuclei evaluated now agree with experiment, where available.

We again stress that the {\it ab initio} NCSM Hamiltonian and the fit
Hamiltonians are pure 2-body Hamiltonians describing the interactions
of all $A=48$ nucleons.  There is no division into valence and core
subsystems, no explicit mean field and no single-particle energies.
In addition, all results that we present are free of effects from
spurious CM motion.  We choose to distinguish the {\it ab initio}
NCSM from the model
with additive phenomenological terms by referring to the latter as simply
the "no-core shell model" or "NCSM" without the "{\it ab initio}" adjective.
For convenience in labeling
figures and tables, we employ $H_{eff}$ or "CD-Bonn" for the former and
"CD-Bonn + 3 terms" for the latter.

Fig. 2 presents the $0\hbar\Omega$ model space results for $^{48}Ca$
with the CD-Bonn effective Hamiltonian, $[H_{eff}]$, at three
values of the basis space parameter, $\hbar\Omega$ = 10 (two cases),
11, and  12 MeV.  For $\hbar\Omega = 10$ MeV, we present results
for the 1995 version of CD-Bonn (column labeled "CDB")
and for the 2000 version (adjacent column labeled "CDB2K") \cite {Machl}
in order to display the minor differences in the spectra from these
two potentials.  

\begin{figure}
\label{fig:Ca48_CDBonn}
\includegraphics[width=\textwidth,clip]{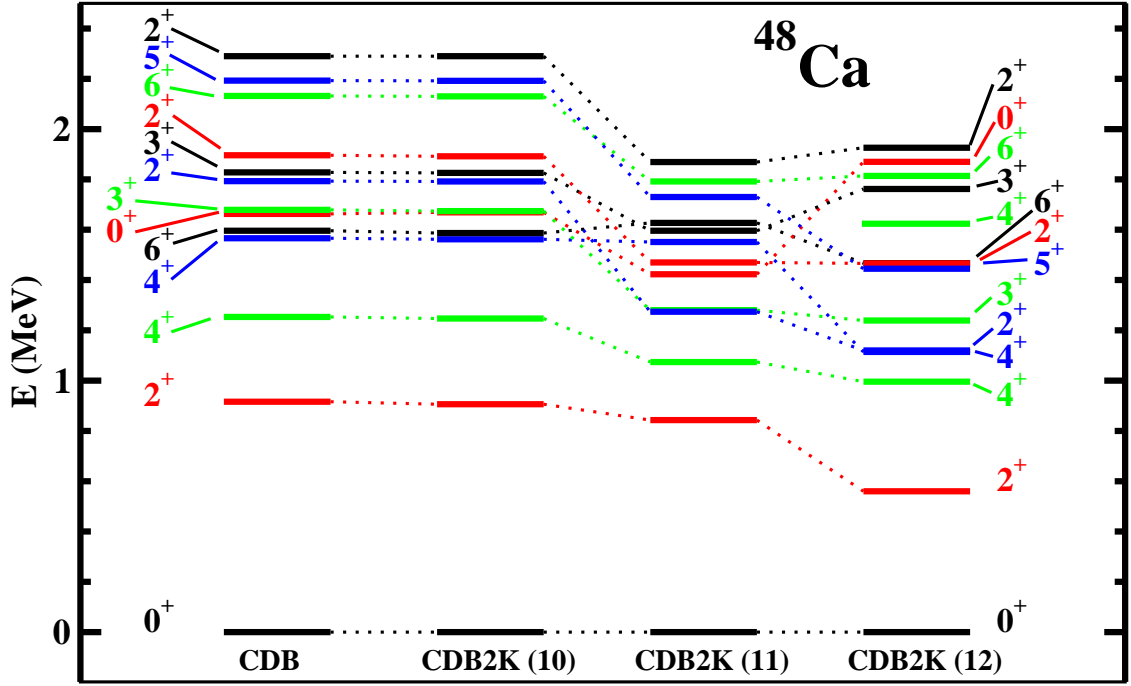}
\vspace{-1.5cm}
\caption{(Color online)  $^{48}Ca$  positive parity excitation
spectra (in MeV) for the CD Bonn (1995) (column 1) effective Hamiltonian
(labeled "CDB") and the CD Bonn (2000) (columns 2-4) effective Hamiltonian
(labeled "CDB2K") in the $0\hbar\Omega$ harmonic oscillator basis space with
$\hbar\Omega$ = 10, 11, and 12 MeV as indicated by the number in
parenthesis in each label. The experimental spectrum is not shown
due to the absence of any experimental states below 3.8 MeV of excitation.
CD-Bonn interactions are taken from Ref. \cite{Machl}. Dotted lines connect nearby states
of the same spin in ascending order.}
\end{figure}

We note that these spectra are very compressed relative to
exeriment -  the first excited state of $^{48}Ca$ is a $2^+$ at 3.832 MeV 
of excitation.  Inspecting the corresponding ground state wavefunctions
reveals an absence of the expected dominance by the
$[0f_{7/2}]^8$ neutron configuration. Instead,
the $1p_{3/2}$ neutron state is significantly populated.
We conclude that the expected energy spacing
between the $0f_{7/2}$ and the
$1p_{3/2}$ state is not supported by the {\it ab initio} NCSM
in such a small model space. This means there is insufficient
spin-orbit splitting.

Here, we can remark that a mean-field study with similarly
derived $H_{eff}$ showed some deficiency in the spin-orbit
splitting for $^{16}O$ in smaller model spaces when
compared with experiment \cite{Hasan}.  In addition, a detailed
study of $^{12}C$ neutrino scattering and magnetic transition
observables revealed that the $a=2$ cluster approximation underpredicts
the spin-orbit splitting needed to explain these data \cite{3body_a}.  This
deficiency appears to be solved by the addition of true NNN
forces and the use of larger basis spaces which are 
beyond the scope of our present efforts.
Detailed investigations of the binding energies and spectroscopy
of p-shell nuclei have also provided strong evidence of the need for
true NNN forces when high-quality NN interactions,
such as CD-Bonn, Argonne V18 or newer chiral interactions, 
are employed \cite{3body_a,3body_b,Nogga06,Navratil07}.  This need is
manifested significantly in the spin-orbit properties.

We do not display the theoretical negative parity states
in this case but we comment that they are similarly compressed
relative to the experimental spreading.  Furthermore,
the lowest negative parity state appears at a rather high excitation energy.
This feature is reminiscent of the results obtained for $^{12}C$
where the negative parity spectra appeared high relative
to the positive parity spectra for model spaces up through
$3 \hbar\Omega$ \cite{NCSM12}.

Note, however, that when we adopt the new strategy
discussed above, using the $H_{eff}$ of the $1 \hbar\Omega$ model
space for both the positive and negative parity states
at $\hbar\Omega = 10 MeV$, the relative spacings of the states
within a given parity are essentially unchanged while
the lowest negative parity excitation above the ground state is
now at 6.9 MeV of excitation, a major improvement.   This will
be discussed shortly along with other results in Fig. 3.

Figs. 3 - 5 display, in column 2 labeled "CDB + 3 terms",
the spectra for these $A=48$ nuclei resulting from our best fit
Hamiltonian as described in the previous section.
The resulting BE/A were presented in Fig. 1 for eight
$A=48$ nuclei with this same Hamiltonian. 
In each of the Figs. 3 - 5, we display the experimental
spectrum in column 1 and the results of Ref. \cite{[CA94]} in column 3.

\begin{figure}[tbhpf]
\label{fig:Ca48_CDBonn+Caurier}
\includegraphics[width=\textwidth,clip]{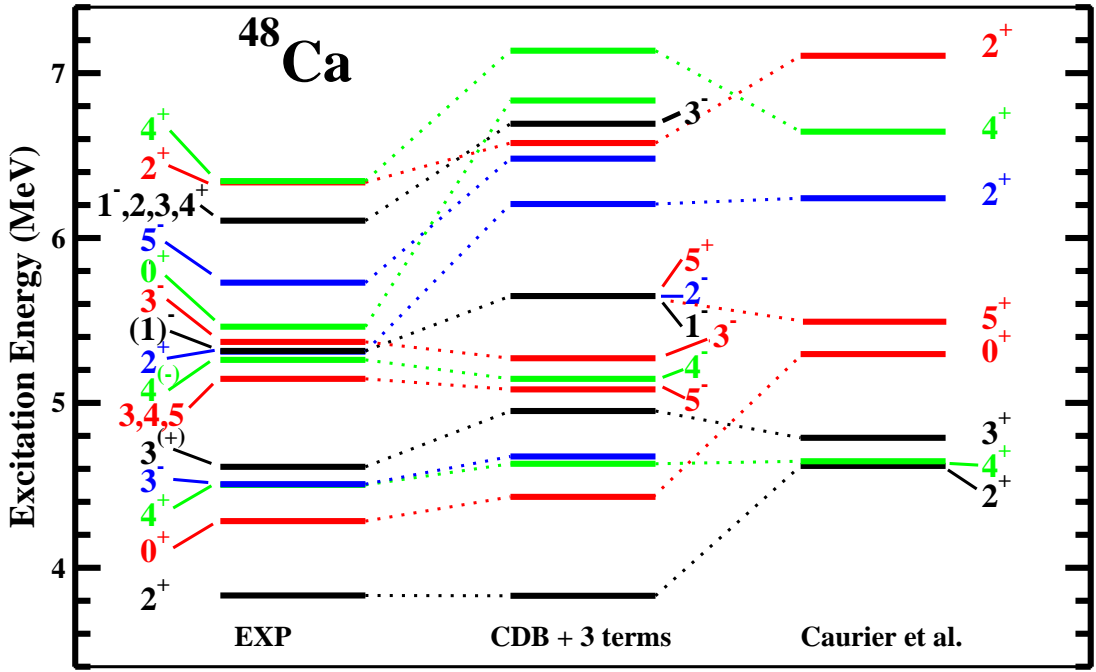}
\caption{(Color online)  $^{48}Ca$ excitation spectra
(in MeV) in the $\hbar\Omega$ = 10 MeV harmonic oscillator basis
for the NCSM effective Hamiltonian plus additive corrections
(column labeled "CDB + 3 terms") compared with experiment and with the results of \cite{[CA94]}
(column labeled "Caurier, et al.").
Positive and negative parity spectra are shown in the first two
columns and the strength parameters are given in the text. Note that
the $0^+$ ground state is omitted from each spectrum and the
theory spectra are shifted so that the ground states all coincide
with the experimental ground state.  Dotted lines connect nearby states
that are likely to be related in character.}
\end{figure}

\begin{figure}[tbhpf]
\label{fig:Sc48_CDBonn+Caurier}
\includegraphics[width=\textwidth]{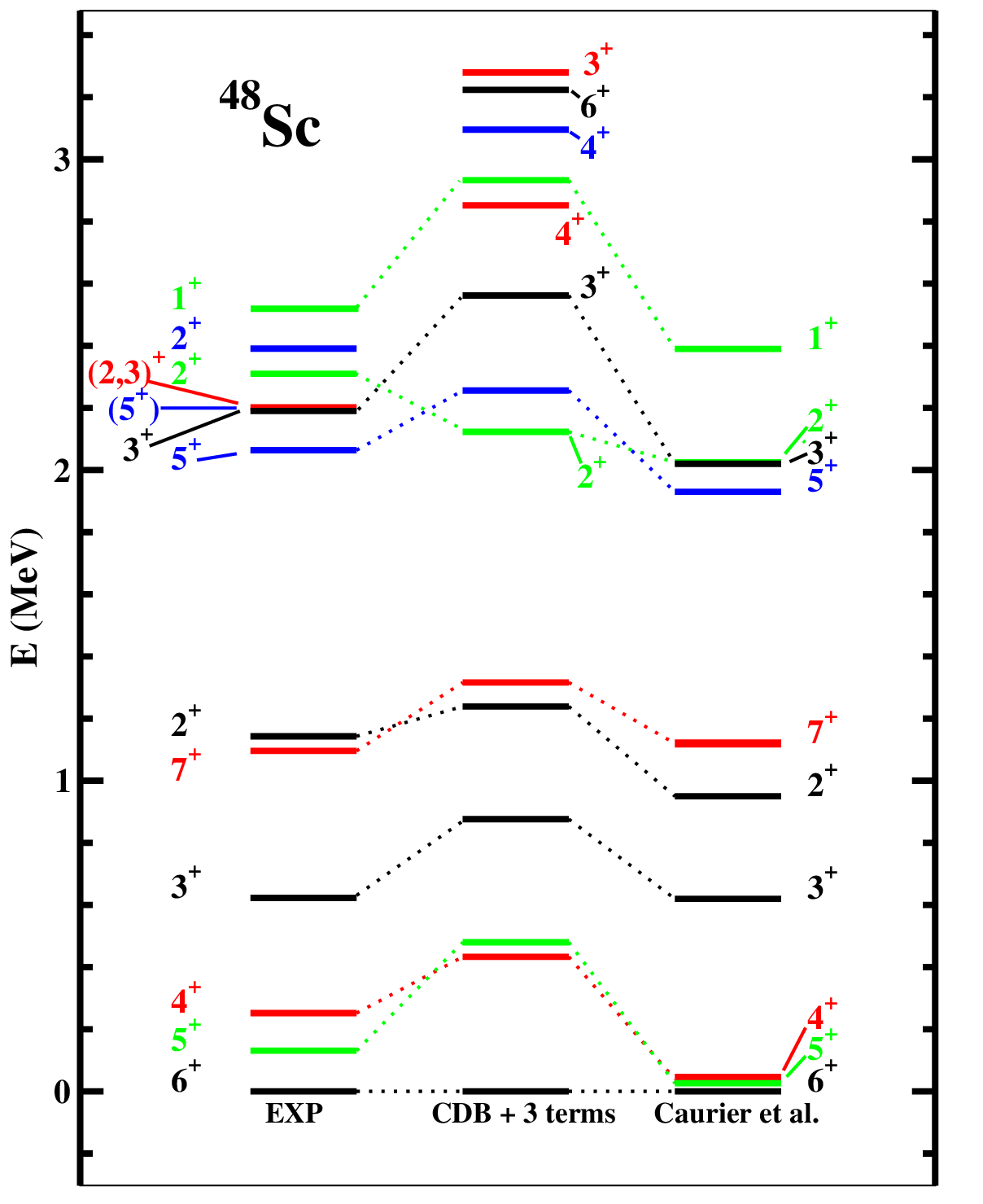}
\caption{(Color online)  $^{48}Sc$ excitation spectra
(in MeV) in the $\hbar\Omega$ = 10 MeV harmonic oscillator basis
for the NCSM effective Hamiltonian plus additive corrections
(column labeled "CDB + 3 terms") compared with experiment and with the results of \cite{[CA94]}
(column labeled "Caurier, et al."). Dotted lines connect nearby states
that are likely to be related in character.}
\end{figure}

\begin{figure}[tbhpf]
\label{fig:Ti48_CDBonn+Caurier}
\includegraphics[height=20cm]{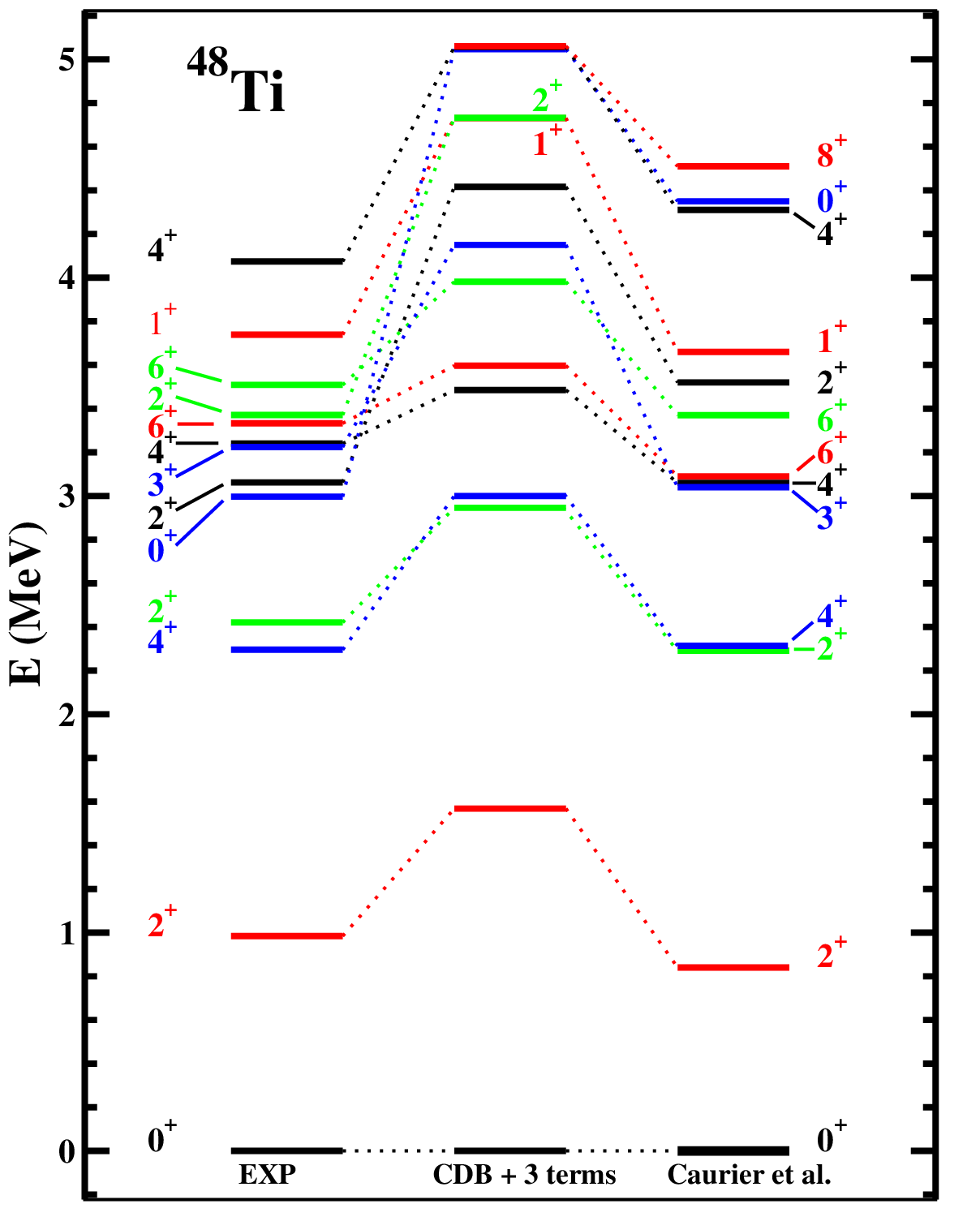}
\caption{(Color online)  $^{48}Ti$ excitation spectra
(in MeV) in the $\hbar\Omega$ = 10 MeV harmonic oscillator basis
for the NCSM effective Hamiltonian plus additive corrections
(column labeled "CDB + 3 terms") compared with experiment and with the results of \cite{[CA94]}
(column labeled "Caurier, et al."). Dotted lines connect nearby states
that are likely to be related in character.}
\end{figure}

As seen in Fig. 3, our fit yields a good description of
the low-lying positive and negative parity states of $^{48}Ca$.
In particular, we observe that the calculated first excited $0^+$,
which was not involved in the fit, appears to be rather close
to the experimental first excited $0^+$.
On the other hand, we are missing another low-lying excited $0^+$.
This may indicate that the intruder state inferred from the
results presented in Ref. \cite{[CA94]} is significantly mixed
between the two low-lying excited $0^+$ states in $^{48}Ca$.

The reasonable agreement of our $^{48}Ca$ negative parity spectrum
with experiment is significant considering that only the position
of the first $3^-$ state was involved in our fit.  It is also
significant since the
negative parity spectrum is sensitive to a set of 2-body
matrix elements that is considerably larger than the set
controlling the positive parity spectrum.  In particular,
we are sensitive to matrix elements involving excitations
from the sd states to the pf states as well as from the
pf states to the sdg states.

Turning now to $^{48}Sc$ shown in Fig. 4 we obtain
one of the more important signatures of the success
of the 3 term fit to these nuclei.  Column 2 of
Fig. 4 shows that we now obtain the correct ground state spin
and a reasonable low-lying positive parity spectrum.  Our spectrum
is slightly more spread than the results of Caurier, et al.,
\cite{[CA94]}, but when comparing Fig. 3 and Fig. 4, please note 
the greatly expanded scale in Fig. 4. 

Our plan to apply these results for $\beta\beta$
processes, leads us to comment on the $1^+$ states in
$^{48}Sc$. There are eight established $1^+$ states below 5.1 MeV
of excitation energy and our 3 term fit spectrum provides only
five $1^+$ states over the same span.  Given our limited
model space and the possibility of intruder states, we may
expect additional states to appear when the model space is
eventually enlarged.  In the meantime, our beta
transition strength function will likely be distributed over a
more limited set of states in a way that will
approximate the distribution among the more dense
experimental spectrum of $1^+$ states.  Of course,
we will include an even larger set of $1^+$ states up to
higher excitation energies when evaluating
the double beta decay rate in a later effort.

Finally, we consider the case of $^{46}Ti$ shown in Fig. 5.
Here again, the low-lying positive parity spectrum from CDB + 3 terms
is in reasonable agreement with experiment except that it is more spread.
We note one particular deficiency -
the $0^+~-~2^+~-~4^+$ theoretical splitting is nearly
that of a vibrator while the experimental spacings indicate
a tendency toward rotational character. The fit of Caurier, et al.,
\cite{[CA94]} succeeds better in this collective property.
It will be interesting
to discover whether the rotational character emerges from
our model as we proceed further into the open shell situation.

We summarize the comparison between our theoretical spectra and 
the experimental spectra for three nuclei involved in the fit in Table
\ref{Aeq48}.  The rms energy deviations between theory and experiment
(excluding states involved in the fit)
indicate that there is considerable room for improvement in the 
spectrum of $^{48}Ti$ and that future fits should include a representative
excited state from this spectrum.

\begin{table}
{\center
\begin{tabular}{|c|c|c|c|}
\hline
  Nucleus/property & $J^{\pi}$ & Exp  & CDBonn~+~3~terms \\ 
\hline
$^{48}Ca:~BE   $ [MeV]  & $0^+$         &  415.991 &  415.948*   \\
$E_{ex}$  [MeV]               & $2^+$         &    3.832    &  3.830*      \\
                                            & $0^+$         &    4.283    &  4.430        \\
                                            & $4^+$         &    4.503    &  4.631        \\
                                            & $3^-$          &    4.675*  &  4.675*      \\
                                       & $3^{(+)}$         &    4.612    &  4.951 $[3^+]$    \\
                                          & $3,4,5$         &    5.146    &  5.082 $[5^-]$     \\
                                       & $4^{(-)}$          &    5.261    &  5.146 $[4^-]$     \\
                                            & $2^+$         &    5.311    &  6.206       \\
                                          & $(1)^-$         &    5.325    &  5.648  $[1^-]$     \\
                                            & $3^-$          &    5.370    &  5.271       \\
$ rms(Exp-Th) $ [MeV]    & -                   &   -              &  0.368       \\
\hline

$^{48}Sc:~BE   $ [MeV]  & $6^+$         &  415.490  &  415.182*   \\
$E_{ex}$  [MeV]               & $5^+$         &      0.131   &      0.480    \\
                                            & $4^+$         &      0.252   &      0.433    \\
                                            & $3^+$         &      0.623   &      0.876    \\
                                            & $7^+$         &      1.096   &      1.316    \\
                                            & $2^+$         &      1.143   &      1.239    \\
                                            & $5^+$         &      2.064   &     2.256    \\
                                            & $3^+$         &      2.190   &     2.562    \\
                                            & $2^+$         &      2.276   &     2.123    \\
                                            & $1^+$         &      2.517   &     2.933    \\
$ rms(Exp-Th) $ [MeV]    & -                   &   -               &     0.268    \\
\hline

$^{48}Ti:~BE   $ [MeV]  & $0^+$         &  418.699   &  418.882*   \\
$E_{ex}$  [MeV]               & $2^+$         &    0.984    &  1.568     \\
                                            & $4^+$         &    2.296    &  3.000     \\
                                            & $2^+$         &    2.421    &  2.946     \\
                                            & $0^+$         &    2.997    &  5.047     \\
                                            & $2^+$         &    3.062    &  4.416     \\
                                            & $3^+$         &    3.224    &  4.150     \\
                                            & $4^+$         &    3.240    &  3.485     \\
                                            & $6^+$         &    3.333    &  3.597     \\
$ rms(Exp-Th) $ [MeV]    & -                   &   -              &  1.008       \\
\hline
\hline
Energy rms (3 nuclei) [MeV]          &   -               & -       &  0.628  \\
\hline
\end{tabular} }
\caption{Binding energies and excitation energies of 
three nuclei from experiment and theory. 
The states indicated with the asterisk are used in the fit
that determines the parameters of the three additive terms 
to the CDBonn effective Hamiltonian. The rms deviations between experiment
and theory are quoted for the listed excited state energies whose spin-parity assignments are reasonably well-established and that are not used in the fitting procedure (8-9 states for each nucleus).  Spin-parity assignments used to relate theory with experiment, when the experimental assignments are uncertain, are indicated in square brackets next to the theoretical state.
The overall energy rms for a total of 25 excited states is quoted at the end of the table.}
\label{Aeq48}
\end{table}

\section{Conclusions and outlook}

Our main goals have been to present the first NCSM results for
$^{48}Ca$, $^{48}Sc$ and $^{48}Ti$ with effective Hamiltonians
derived directly from a realistic NN interaction and to investigate
phenomenological improvements.  The {\it ab initio} NCSM results
display the shortcomings of the limited model spaces presently
available as well as possible shortcomings from neglecting
three-body forces.  We answer the question of
whether the NCSM can be adjusted to obtain reasonable
fits with additive phenomenological two-body potentials in the
affirmative.  In particular, we show that additive isospin-dependent
central terms plus a tensor force can achieve accurate
BE/A and reasonable spectra for these three systems.
In addition, accurate BE/A are obtained for
eight $A=48$ nuclei reproducing the experimental valley of
stability. The net change of interaction energies
is of the order of a few percent with the added phenomenological
terms. More extensive searching could undoubtedly improve the
fits to the low-lying spectra.

Future efforts motivated by the present results are many-fold.
We intend to improve the {\it ab initio} $H_{eff}$ by extensions to the
three-body cluster approximation and to include three-nucleon
interactions.  We forsee initial applications to $\beta\beta$ decay,
both the $\nu\nu$ and $0\nu$ decay channels, by first extending our
calculations to the Gamow-Teller (+/-) strengths.  In the near future,
we will be able to address significantly larger basis states as well.
As a first step, we explicitly calculated the Ikeda sum rule and
found that it is obeyed exactly in the NCSM.

These future efforts represent major undertakings and depend on
major increases in computational resources.  We firmly believe these
planned endeavors are warranted in light of the importance of retaining
as much predictive power as possible when addressing
$\beta\beta$ decay.

\section{Acknowledgements}
We thank Vesselin Gueorguiev, Christian Forssen and Mihai Horoi 
for useful discussions.
This work was partly performed under the auspices of
the U. S. Department of Energy by the University of California,
Lawrence Livermore National Laboratory under contract
No. W-7405-Eng-48.
This work was also supported in part by USDOE grant DE-FG-02 87ER40371,
Division of Nuclear Physics.   This work was also supported in part by
NSF grant INT0070789.

\pagebreak

\end{document}